\documentclass{svmult}
\usepackage{epsfig,graphicx} 
\usepackage[bottom]{footmisc}
\usepackage{natbib,url}      
\bibpunct{(}{)}{;}{a}{}{,}   
\def\cite#1{\citealp{#1}}    

\def\authorindex#1{}

\begin{document}\newcount\preprintheader\preprintheader=1



\def\thisvolume{these proceedings}

\def\aj{{AJ}}			
\def\araa{{ARA\&A}}		
\def\apj{{ApJ}}			
\def\apjl{{ApJ}}		
\def\apjs{{ApJS}}		
\def\ao{{Appl.\ Optics}} 
\def\apss{{Ap\&SS}}		
\def\aap{{A\&A}}		
\def\aapr{{A\&A~Rev.}}		
\def\aaps{{A\&AS}}		
\def\aspcs{{ASP Conf.\ Ser.}}
\def\azh{{AZh}}			
\def\baas{{BAAS}}		
\def\jrasc{{JRASC}}		
\def\memras{{MmRAS}}		
\def\mnras{{MNRAS}}
\def\nat{{Nat}}		
\def\pra{{Phys.\ Rev.\ A}} 
\def\prb{{Phys.\ Rev.\ B}}		
\def\prc{{Phys.\ Rev.\ C}}		
\def\prd{{Phys.\ Rev.\ D}}		
\def\prl{{Phys.\ Rev.\ Lett}}	
\def\pasp{{PASP}}
\def\pasj{{PASJ}}		
\def\qjras{{QJRAS}}
\def\science{{Sci}}		
\def\skytel{{S\&T}}		
\def\solphys{{Solar\ Phys.}} 
\def\sovast{{Soviet\ Ast.}}  
\def\ssr{{Space\ Sci.\ Rev.}}
\def\svassp{{Astrophys.\ Space Science Proc.}}
\def\zap{{ZAp}}			
\let\astap=\aap
\let\apjlett=\apjl
\let\apjsupp=\apjs

\def\ion#1#2{{\rm #1}\,{\uppercase{#2}}}  
\def\deg{\hbox{$^\circ$}}
\def\sun{\hbox{$\odot$}}
\def\earth{\hbox{$\oplus$}}
\def\la{\mathrel{\hbox{\rlap{\hbox{\lower4pt\hbox{$\sim$}}}\hbox{$<$}}}}
\def\ga{\mathrel{\hbox{\rlap{\hbox{\lower4pt\hbox{$\sim$}}}\hbox{$>$}}}}
\def\sq{\hbox{\rlap{$\sqcap$}$\sqcup$}}
\def\arcmin{\hbox{$^\prime$}}
\def\arcsec{\hbox{$^{\prime\prime}$}}
\def\fd{\hbox{$.\!\!^{\rm d}$}}
\def\fh{\hbox{$.\!\!^{\rm h}$}}
\def\fm{\hbox{$.\!\!^{\rm m}$}}
\def\fs{\hbox{$.\!\!^{\rm s}$}}
\def\fdg{\hbox{$.\!\!^\circ$}}
\def\farcm{\hbox{$.\mkern-4mu^\prime$}}
\def\farcs{\hbox{$.\!\!^{\prime\prime}$}}
\def\fp{\hbox{$.\!\!^{\scriptscriptstyle\rm p}$}}
\def\micron{\hbox{$\mu$m}}
\def\onehalf{\hbox{$\,^1\!/_2$}}	
\def\onethird{\hbox{$\,^1\!/_3$}}
\def\twothirds{\hbox{$\,^2\!/_3$}}
\def\onequarter{\hbox{$\,^1\!/_4$}}
\def\threequarters{\hbox{$\,^3\!/_4$}}
\def\ubv{\hbox{$U\!BV$}}		
\def\ubvr{\hbox{$U\!BV\!R$}}		
\def\ubvri{\hbox{$U\!BV\!RI$}}		
\def\ubvrij{\hbox{$U\!BV\!RI\!J$}}		
\def\ubvrijh{\hbox{$U\!BV\!RI\!J\!H$}}		
\def\ubvrijhk{\hbox{$U\!BV\!RI\!J\!H\!K$}}		
\def\ub{\hbox{$U\!-\!B$}}		
\def\bv{\hbox{$B\!-\!V$}}		
\def\vr{\hbox{$V\!-\!R$}}		
\def\ur{\hbox{$U\!-\!R$}}


\def\labelitemi{{\bf --}}  

\def\rmit#1{{\it #1}}              
\def\rmit#1{{\rm #1}}              
\def\etal{\rmit{et al.}}           
\def\etc{\rmit{etc.}}           
\def\ie{\rmit{i.e.,}}              
\def\eg{\rmit{e.g.,}}              
\def\cf{cf.}                       
\def\viz{\rmit{viz.}}
\def\vs{\rmit{vs.}}

\def\rot{\hbox{\rm rot}}
\def\div{\hbox{\rm div}}
\def\lesssim{\mathrel{\hbox{\rlap{\hbox{\lower4pt\hbox{$\sim$}}}\hbox{$<$}}}}
\def\gtrsim{\mathrel{\hbox{\rlap{\hbox{\lower4pt\hbox{$\sim$}}}\hbox{$>$}}}}
\def\dif{\: {\rm d}}                       
\def\ep{\:{\rm e}^}                        
\def\dash{\hbox{$\,-\,$}}                  
\def\is{\!=\!}                             

\def\starname#1#2{${#1}$\,{\rm {#2}}}  
\def\Teff{\hbox{$T_{\rm eff}$}}   

\def\kms{\hbox{km$\;$s$^{-1}$}}
\def\Mxcm{\hbox{Mx\,cm$^{-2}$}}    

\def\Bapp{\hbox{$B_{\rm app}$}}    

\def\komega{($k, \omega$)}                 
\def\kf{($k_h,f$)}                         
\def\VminI{\hbox{$V\!\!-\!\!I$}}           
\def\IminI{\hbox{$I\!\!-\!\!I$}}           
\def\VminV{\hbox{$V\!\!-\!\!V$}}           
\def\Xt{\hbox{$X\!\!-\!t$}}                

\def\level #1 #2#3#4{$#1 \: ^{#2} \mbox{#3} ^{#4}$}   

\def\specchar#1{\uppercase{#1}}    
\def\AlI{\mbox{Al\,\specchar{i}}}  
\def\BI{\mbox{B\,\specchar{i}}} 
\def\BII{\mbox{B\,\specchar{ii}}}  
\def\BaI{\mbox{Ba\,\specchar{i}}}  
\def\BaII{\mbox{Ba\,\specchar{ii}}} 
\def\CI{\mbox{C\,\specchar{i}}} 
\def\CII{\mbox{C\,\specchar{ii}}} 
\def\CIII{\mbox{C\,\specchar{iii}}} 
\def\CIV{\mbox{C\,\specchar{iv}}} 
\def\CaI{\mbox{Ca\,\specchar{i}}} 
\def\CaII{\mbox{Ca\,\specchar{ii}}} 
\def\CaIII{\mbox{Ca\,\specchar{iii}}} 
\def\CoI{\mbox{Co\,\specchar{i}}} 
\def\CrI{\mbox{Cr\,\specchar{i}}} 
\def\CriI{\mbox{Cr\,\specchar{ii}}} 
\def\CsI{\mbox{Cs\,\specchar{i}}} 
\def\CsII{\mbox{Cs\,\specchar{ii}}} 
\def\CuI{\mbox{Cu\,\specchar{i}}} 
\def\FeI{\mbox{Fe\,\specchar{i}}} 
\def\FeII{\mbox{Fe\,\specchar{ii}}} 
\def\FeIX{\mbox{Fe\,\specchar{ix}}}
\def\FeX{\mbox{Fe\,\specchar{x}}}
\def\FeXVI{\mbox{Fe\,\specchar{xvi}}}
\def\FrI{\mbox{Fr\,\specchar{i}}}
\def\HI{\mbox{H\,\specchar{i}}} 
\def\HII{\mbox{H\,\specchar{ii}}} 
\def\Hmin{\hbox{\rmH$^{^{_{\scriptstyle -}}}$}}      
\def\Hemin{\hbox{{\rm He}$^{^{_{\scriptstyle -}}}$}} 
\def\HeI{\mbox{He\,\specchar{i}}} 
\def\HeII{\mbox{He\,\specchar{ii}}} 
\def\HeIII{\mbox{He\,\specchar{iii}}} 
\def\KI{\mbox{K\,\specchar{i}}} 
\def\KII{\mbox{K\,\specchar{ii}}} 
\def\KIII{\mbox{K\,\specchar{iii}}} 
\def\LiI{\mbox{Li\,\specchar{i}}} 
\def\LiII{\mbox{Li\,\specchar{ii}}} 
\def\LiIII{\mbox{Li\,\specchar{iii}}} 
\def\MgI{\mbox{Mg\,\specchar{i}}} 
\def\MgII{\mbox{Mg\,\specchar{ii}}} 
\def\MgIII{\mbox{Mg\,\specchar{iii}}} 
\def\MnI{\mbox{Mn\,\specchar{i}}} 
\def\NI{\mbox{N\,\specchar{i}}}
\def\NaI{\mbox{Na\,\specchar{i}}}
\def\NaII{\mbox{Na\,\specchar{ii}}}
\def\NaIII{\mbox{Na\,\specchar{iii}}} 
\def\NiI{\mbox{Ni\,\specchar{i}}} 
\def\NiII{\mbox{Ni\,\specchar{ii}}}
\def\NiIII{\mbox{Ni\,\specchar{iii}}} 
\def\OI{\mbox{O\,\specchar{i}}} 
\def\OVI{\mbox{O\,\specchar{vi}}}
\def\RbI{\mbox{Rb\,\specchar{i}}} 
\def\SII{\mbox{S\,\specchar{ii}}} 
\def\SiI{\mbox{Si\,\specchar{i}}} 
\def\SiII{\mbox{Si\,\specchar{ii}}} 
\def\SrI{\mbox{Sr\,\specchar{i}}}
\def\SrII{\mbox{Sr\,\specchar{ii}}}
\def\TiI{\mbox{Ti\,\specchar{i}}} 
\def\TiII{\mbox{Ti\,\specchar{ii}}} 
\def\TiIII{\mbox{Ti\,\specchar{iii}}} 
\def\TiIV{\mbox{Ti\,\specchar{iv}}} 
\def\VI{\mbox{V\,\specchar{i}}} 
\def\HtwoO{\mbox{H$_2$O}}        
\def\Otwo{\mbox{O$_2$}}          

\def\Halpha{\mbox{H\hspace{0.1ex}$\alpha$}} 
\def\Ha{\mbox{H\hspace{0.2ex}$\alpha$}}
\def\Hbeta{\mbox{H\hspace{0.2ex}$\beta$}}
\def\Hgamma{\mbox{H\hspace{0.2ex}$\gamma$}}
\def\Hdelta{\mbox{H\hspace{0.2ex}$\delta$}}
\def\Hepsilon{\mbox{H\hspace{0.2ex}$\epsilon$}}
\def\Hzeta{\mbox{H\hspace{0.2ex}$\zeta$}}
\def\Lyalpha{\mbox{Ly$\hspace{0.2ex}\alpha$}}
\def\Lybeta{\mbox{Ly$\hspace{0.2ex}\beta$}}
\def\Lygamma{\mbox{Ly$\hspace{0.2ex}\gamma$}}
\def\Lycont{\mbox{Ly\hspace{0.2ex}{\small cont}}}
\def\Baalpha{\mbox{Ba$\hspace{0.2ex}\alpha$}}
\def\Babeta{\mbox{Ba$\hspace{0.2ex}\beta$}}
\def\Bacont{\mbox{Ba\hspace{0.2ex}{\small cont}}}
\def\Paalpha{\mbox{Pa$\hspace{0.2ex}\alpha$}}
\def\Bralpha{\mbox{Br$\hspace{0.2ex}\alpha$}}

\def\NaD{\mbox{Na\,\specchar{i}\,D}}    
\def\NaDone{\mbox{Na\,\specchar{i}\,\,D$_1$}}
\def\NaDtwo{\mbox{Na\,\specchar{i}\,\,D$_2$}}
\def\NaID{\mbox{Na\,\specchar{i}\,\,D}}
\def\NaIDone{\mbox{Na\,\specchar{i}\,\,D$_1$}}
\def\NaIDtwo{\mbox{Na\,\specchar{i}\,\,D$_2$}}
\def\Done{\mbox{D$_1$}}
\def\Dtwo{\mbox{D$_2$}}

\def\Mgbone{\mbox{Mg\,\specchar{i}\,b$_1$}}
\def\Mgbtwo{\mbox{Mg\,\specchar{i}\,b$_2$}}
\def\Mgbthree{\mbox{Mg\,\specchar{i}\,b$_3$}}
\def\MgIb{\mbox{Mg\,\specchar{i}\,b}}
\def\MgIbone{\mbox{Mg\,\specchar{i}\,b$_1$}}
\def\MgIbtwo{\mbox{Mg\,\specchar{i}\,b$_2$}}
\def\MgIbthree{\mbox{Mg\,\specchar{i}\,b$_3$}}

\def\CaIIK{\mbox{Ca\,\specchar{ii}\,K}}       
\def\CaIIH{\mbox{Ca\,\specchar{ii}\,H}}
\def\CaIIHK{\mbox{Ca\,\specchar{ii}\,H\,\&\,K}}
\def\HK{\mbox{H\,\&\,K}}
\def\Kthree{\mbox{K$_3$}}      
\def\Hthree{\mbox{H$_3$}}
\def\Ktwo{\mbox{K$_2$}}
\def\Htwo{\mbox{H$_2$}}
\def\Kone{\mbox{K$_1$}}     
\def\Hone{\mbox{H$_1$}}     
\def\KtwoV{\mbox{K$_{2V}$}}
\def\KtwoR{\mbox{K$_{2R}$}}
\def\KoneV{\mbox{K$_{1V}$}}
\def\KoneR{\mbox{K$_{1R}$}}
\def\HtwoV{\mbox{H$_{2V}$}}
\def\HtwoR{\mbox{H$_{2R}$}}
\def\HoneV{\mbox{H$_{1V}$}}
\def\HoneR{\mbox{H$_{1R}$}}

\def\hk{\mbox{h\,\&\,k}}
\def\kthree{\mbox{k$_3$}}    
\def\hthree{\mbox{h$_3$}}
\def\ktwo{\mbox{k$_2$}}
\def\htwo{\mbox{h$_2$}}
\def\kone{\mbox{k$_1$}}     
\def\hone{\mbox{h$_1$}}     
\def\ktwoV{\mbox{k$_{2V}$}}
\def\ktwoR{\mbox{k$_{2R}$}}
\def\koneV{\mbox{k$_{1V}$}}
\def\koneR{\mbox{k$_{1R}$}}
\def\htwoV{\mbox{h$_{2V}$}}
\def\htwoR{\mbox{h$_{2R}$}}
\def\honeV{\mbox{h$_{1V}$}}
\def\honeR{\mbox{h$_{1R}$}}


\def\thisvolume{these proceedings}

\def\aj{{AJ}}			
\def\araa{{ARA\&A}}		
\def\apj{{ApJ}}			
\def\apjl{{ApJ}}		
\def\apjs{{ApJS}}		
\def\ao{{Appl.\ Optics}} 
\def\apss{{Ap\&SS}}		
\def\aap{{A\&A}}		
\def\aapr{{A\&A~Rev.}}		
\def\aaps{{A\&AS}}		
\def\an{{Astron.\ Nachrichten}}
\def\aspcs{{ASP Conf.\ Ser.}}
\def\assp{{Astrophys.\ \& Space Sci.\ Procs., Springer, Heidelberg}}
\def\azh{{AZh}}			
\def\baas{{BAAS}}		
\def\jrasc{{JRASC}}	
\def\memras{{MmRAS}}		
\def\mnras{{MNRAS}}
\def\nat{{Nat}}		
\def\pra{{Phys.\ Rev.\ A}} 
\def\prb{{Phys.\ Rev.\ B}}		
\def\prc{{Phys.\ Rev.\ C}}		
\def\prd{{Phys.\ Rev.\ D}}		
\def\prl{{Phys.\ Rev.\ Lett.}} 
\def\pasp{{PASP}}
\def\pasj{{PASJ}}		
\def\qjras{{QJRAS}}
\def\science{{Sci}}		
\def\skytel{{S\&T}}		
\def\solphys{{Solar\ Phys.}} 
\def\sovast{{Soviet\ Ast.}}  
\def\ssr{{Space\ Sci.\ Rev.}}
\def\svassp{{Astrophys.\ Space Sci.\ Procs., Springer, Heidelberg}}
\def\zap{{ZAp}}			
\let\astap=\aap
\let\apjlett=\apjl
\let\apjsupp=\apjs
\def\grl{{Geophys.\ Res.\ Lett.}}  
\def\jgr{{J. Geophys.\ Res.}} 

\def\ion#1#2{{\rm #1}\,{\uppercase{#2}}}  
\def\deg{\hbox{$^\circ$}}
\def\sun{\hbox{$\odot$}}
\def\earth{\hbox{$\oplus$}}
\def\la{\mathrel{\hbox{\rlap{\hbox{\lower4pt\hbox{$\sim$}}}\hbox{$<$}}}}
\def\ga{\mathrel{\hbox{\rlap{\hbox{\lower4pt\hbox{$\sim$}}}\hbox{$>$}}}}
\def\sq{\hbox{\rlap{$\sqcap$}$\sqcup$}}
\def\arcmin{\hbox{$^\prime$}}
\def\arcsec{\hbox{$^{\prime\prime}$}}
\def\fd{\hbox{$.\!\!^{\rm d}$}}
\def\fh{\hbox{$.\!\!^{\rm h}$}}
\def\fm{\hbox{$.\!\!^{\rm m}$}}
\def\fs{\hbox{$.\!\!^{\rm s}$}}
\def\fdg{\hbox{$.\!\!^\circ$}}
\def\farcm{\hbox{$.\mkern-4mu^\prime$}}
\def\farcs{\hbox{$.\!\!^{\prime\prime}$}}
\def\fp{\hbox{$.\!\!^{\scriptscriptstyle\rm p}$}}
\def\micron{\hbox{$\mu$m}}
\def\onehalf{\hbox{$\,^1\!/_2$}}	
\def\onethird{\hbox{$\,^1\!/_3$}}
\def\twothirds{\hbox{$\,^2\!/_3$}}
\def\onequarter{\hbox{$\,^1\!/_4$}}
\def\threequarters{\hbox{$\,^3\!/_4$}}
\def\ubv{\hbox{$U\!BV$}}		
\def\ubvr{\hbox{$U\!BV\!R$}}		
\def\ubvri{\hbox{$U\!BV\!RI$}}		
\def\ubvrij{\hbox{$U\!BV\!RI\!J$}}		
\def\ubvrijh{\hbox{$U\!BV\!RI\!J\!H$}}		
\def\ubvrijhk{\hbox{$U\!BV\!RI\!J\!H\!K$}}		
\def\ub{\hbox{$U\!-\!B$}}		
\def\bv{\hbox{$B\!-\!V$}}		
\def\vr{\hbox{$V\!-\!R$}}		
\def\ur{\hbox{$U\!-\!R$}}


\def\labelitemi{{\bf --}}  

\def\rmit#1{{\it #1}}              
\def\rmit#1{{\rm #1}}              
\def\etal{\rmit{et al.}}           
\def\etc{\rmit{etc.}}           
\def\ie{\rmit{i.e.,}}              
\def\eg{\rmit{e.g.,}}              
\def\cf{cf.}                       
\def\viz{\rmit{viz.}}
\def\vs{\rmit{vs.}}

\def\rot{\hbox{\rm rot}}
\def\div{\hbox{\rm div}}
\def\lesssim{\mathrel{\hbox{\rlap{\hbox{\lower4pt\hbox{$\sim$}}}\hbox{$<$}}}}
\def\gtrsim{\mathrel{\hbox{\rlap{\hbox{\lower4pt\hbox{$\sim$}}}\hbox{$>$}}}}
\def\mathstacksym#1#2#3#4#5{\def#1{\mathrel{\hbox to 0pt{\lower 
    #5\hbox{#3}\hss} \raise #4\hbox{#2}}}}
\mathstacksym\lesssim{$<$}{$\sim$}{1.5pt}{3.5pt} 
\mathstacksym\gtrsim{$>$}{$\sim$}{1.5pt}{3.5pt} 
\mathstacksym\lrarrow{$\leftarrow$}{$\rightarrow$}{2pt}{1pt} 
\mathstacksym\lessgreat{$>$}{$<$}{3pt}{3pt} 

\def\dif{\: {\rm d}}                       
\def\ep{\:{\rm e}^}                        
\def\dash{\hbox{$\,-\,$}}                  
\def\is{\!=\!}                             

\def\starname#1#2{${#1}$\,{\rm {#2}}}  
\def\Teff{\hbox{$T_{\rm eff}$}}   

\def\kms{\hbox{km$\;$s$^{-1}$}}
\def\ms{\hbox{m$\;$s$^{-1}$}}
\def\Mxcm{\hbox{Mx\,cm$^{-2}$}}    

\def\Bapp{\hbox{$B_{\rm app}$}}    

\def\komega{($k, \omega$)}                 
\def\kf{($k_h,f$)}                         
\def\VminI{\hbox{$V\!\!-\!\!I$}}           
\def\IminI{\hbox{$I\!\!-\!\!I$}}           
\def\VminV{\hbox{$V\!\!-\!\!V$}}           
\def\Xt{\hbox{$X\!\!-\!t$}}                

\def\level #1 #2#3#4{$#1 \: ^{#2} \mbox{#3} ^{#4}$}   

\def\specchar#1{\uppercase{#1}}    
\def\AlI{\mbox{Al\,\specchar{i}}}  
\def\BI{\mbox{B\,\specchar{i}}} 
\def\BII{\mbox{B\,\specchar{ii}}}  
\def\BaI{\mbox{Ba\,\specchar{i}}}  
\def\BaII{\mbox{Ba\,\specchar{ii}}} 
\def\CI{\mbox{C\,\specchar{i}}} 
\def\CII{\mbox{C\,\specchar{ii}}} 
\def\CIII{\mbox{C\,\specchar{iii}}} 
\def\CIV{\mbox{C\,\specchar{iv}}} 
\def\CaI{\mbox{Ca\,\specchar{i}}} 
\def\CaII{\mbox{Ca\,\specchar{ii}}} 
\def\CaIII{\mbox{Ca\,\specchar{iii}}} 
\def\CoI{\mbox{Co\,\specchar{i}}} 
\def\CrI{\mbox{Cr\,\specchar{i}}} 
\def\CriI{\mbox{Cr\,\specchar{ii}}} 
\def\CsI{\mbox{Cs\,\specchar{i}}} 
\def\CsII{\mbox{Cs\,\specchar{ii}}} 
\def\CuI{\mbox{Cu\,\specchar{i}}} 
\def\FeI{\mbox{Fe\,\specchar{i}}} 
\def\FeII{\mbox{Fe\,\specchar{ii}}} 
\def\FeIX{\mbox{Fe\,\specchar{ix}}}
\def\FeX{\mbox{Fe\,\specchar{x}}}
\def\FeXVI{\mbox{Fe\,\specchar{xvi}}}
\def\FrI{\mbox{Fr\,\specchar{i}}}
\def\HI{\mbox{H\,\specchar{i}}} 
\def\HII{\mbox{H\,\specchar{ii}}} 
\def\Hmin{\hbox{\rmH$^{^{_{\scriptstyle -}}}$}}      
\def\Hemin{\hbox{{\rm He}$^{^{_{\scriptstyle -}}}$}} 
\def\HeI{\mbox{He\,\specchar{i}}} 
\def\HeII{\mbox{He\,\specchar{ii}}} 
\def\HeIII{\mbox{He\,\specchar{iii}}} 
\def\KI{\mbox{K\,\specchar{i}}} 
\def\KII{\mbox{K\,\specchar{ii}}} 
\def\KIII{\mbox{K\,\specchar{iii}}} 
\def\LiI{\mbox{Li\,\specchar{i}}} 
\def\LiII{\mbox{Li\,\specchar{ii}}} 
\def\LiIII{\mbox{Li\,\specchar{iii}}} 
\def\MgI{\mbox{Mg\,\specchar{i}}} 
\def\MgII{\mbox{Mg\,\specchar{ii}}} 
\def\MgIII{\mbox{Mg\,\specchar{iii}}} 
\def\MnI{\mbox{Mn\,\specchar{i}}} 
\def\NI{\mbox{N\,\specchar{i}}}
\def\NIV{\mbox{N\,\specchar{iv}}}
\def\NaI{\mbox{Na\,\specchar{i}}}
\def\NaII{\mbox{Na\,\specchar{ii}}}
\def\NaIII{\mbox{Na\,\specchar{iii}}}
\def\NeVIII{\mbox{Ne\,\specchar{viii}}} 
\def\NiI{\mbox{Ni\,\specchar{i}}} 
\def\NiII{\mbox{Ni\,\specchar{ii}}}
\def\NiIII{\mbox{Ni\,\specchar{iii}}} 
\def\OI{\mbox{O\,\specchar{i}}} 
\def\OVI{\mbox{O\,\specchar{vi}}}
\def\RbI{\mbox{Rb\,\specchar{i}}} 
\def\SII{\mbox{S\,\specchar{ii}}} 
\def\SiI{\mbox{Si\,\specchar{i}}} 
\def\SiII{\mbox{Si\,\specchar{ii}}} 
\def\SrI{\mbox{Sr\,\specchar{i}}}
\def\SrII{\mbox{Sr\,\specchar{ii}}}
\def\TiI{\mbox{Ti\,\specchar{i}}} 
\def\TiII{\mbox{Ti\,\specchar{ii}}} 
\def\TiIII{\mbox{Ti\,\specchar{iii}}} 
\def\TiIV{\mbox{Ti\,\specchar{iv}}} 
\def\VI{\mbox{V\,\specchar{i}}} 
\def\HtwoO{\mbox{H$_2$O}}        
\def\Otwo{\mbox{O$_2$}}          

\def\Halpha{\mbox{H\hspace{0.1ex}$\alpha$}} 
\def\Ha{\mbox{H\hspace{0.2ex}$\alpha$}}
\def\Hbeta{\mbox{H\hspace{0.2ex}$\beta$}}
\def\Hgamma{\mbox{H\hspace{0.2ex}$\gamma$}}
\def\Hdelta{\mbox{H\hspace{0.2ex}$\delta$}}
\def\Hepsilon{\mbox{H\hspace{0.2ex}$\epsilon$}}
\def\Hzeta{\mbox{H\hspace{0.2ex}$\zeta$}}
\def\Lyalpha{\mbox{Ly$\hspace{0.2ex}\alpha$}}
\def\Lybeta{\mbox{Ly$\hspace{0.2ex}\beta$}}
\def\Lygamma{\mbox{Ly$\hspace{0.2ex}\gamma$}}
\def\Lycont{\mbox{Ly\hspace{0.2ex}{\small cont}}}
\def\Baalpha{\mbox{Ba$\hspace{0.2ex}\alpha$}}
\def\Babeta{\mbox{Ba$\hspace{0.2ex}\beta$}}
\def\Bacont{\mbox{Ba\hspace{0.2ex}{\small cont}}}
\def\Paalpha{\mbox{Pa$\hspace{0.2ex}\alpha$}}
\def\Bralpha{\mbox{Br$\hspace{0.2ex}\alpha$}}

\def\NaD{\mbox{Na\,\specchar{i}\,D}}    
\def\NaDone{\mbox{Na\,\specchar{i}\,\,D$_1$}}
\def\NaDtwo{\mbox{Na\,\specchar{i}\,\,D$_2$}}
\def\NaID{\mbox{Na\,\specchar{i}\,\,D}}
\def\NaIDone{\mbox{Na\,\specchar{i}\,\,D$_1$}}
\def\NaIDtwo{\mbox{Na\,\specchar{i}\,\,D$_2$}}
\def\Done{\mbox{D$_1$}}
\def\Dtwo{\mbox{D$_2$}}

\def\Mgbone{\mbox{Mg\,\specchar{i}\,b$_1$}}
\def\Mgbtwo{\mbox{Mg\,\specchar{i}\,b$_2$}}
\def\Mgbthree{\mbox{Mg\,\specchar{i}\,b$_3$}}
\def\MgIb{\mbox{Mg\,\specchar{i}\,b}}
\def\MgIbone{\mbox{Mg\,\specchar{i}\,b$_1$}}
\def\MgIbtwo{\mbox{Mg\,\specchar{i}\,b$_2$}}
\def\MgIbthree{\mbox{Mg\,\specchar{i}\,b$_3$}}

\def\CaIIK{\mbox{Ca\,\specchar{ii}\,K}}       
\def\CaIIH{\mbox{Ca\,\specchar{ii}\,H}}
\def\CaIIHK{\mbox{Ca\,\specchar{ii}\,H\,\&\,K}}
\def\HK{\mbox{H\,\&\,K}}
\def\Kthree{\mbox{K$_3$}}      
\def\Hthree{\mbox{H$_3$}}
\def\Ktwo{\mbox{K$_2$}}
\def\Htwo{\mbox{H$_2$}}
\def\Kone{\mbox{K$_1$}}     
\def\Hone{\mbox{H$_1$}}     
\def\KtwoV{\mbox{K$_{2V}$}}
\def\KtwoR{\mbox{K$_{2R}$}}
\def\KoneV{\mbox{K$_{1V}$}}
\def\KoneR{\mbox{K$_{1R}$}}
\def\HtwoV{\mbox{H$_{2V}$}}
\def\HtwoR{\mbox{H$_{2R}$}}
\def\HoneV{\mbox{H$_{1V}$}}
\def\HoneR{\mbox{H$_{1R}$}}

\def\hk{\mbox{h\,\&\,k}}
\def\kthree{\mbox{k$_3$}}    
\def\hthree{\mbox{h$_3$}}
\def\ktwo{\mbox{k$_2$}}
\def\htwo{\mbox{h$_2$}}
\def\kone{\mbox{k$_1$}}     
\def\hone{\mbox{h$_1$}}     
\def\ktwoV{\mbox{k$_{2V}$}}
\def\ktwoR{\mbox{k$_{2R}$}}
\def\koneV{\mbox{k$_{1V}$}}
\def\koneR{\mbox{k$_{1R}$}}
\def\htwoV{\mbox{h$_{2V}$}}
\def\htwoR{\mbox{h$_{2R}$}}
\def\honeV{\mbox{h$_{1V}$}}
\def\honeR{\mbox{h$_{1R}$}}

\ifnum\preprintheader=1     
\makeatletter  
\def\@maketitle{\newpage
\markboth{}{}%
  {\mbox{} \vspace*{-8ex} \par 
   \em \footnotesize To appear in ``Magnetic Coupling between the Interior 
       and the Atmosphere of the Sun'', eds. S.~S.~Hasan and R.~J.~Rutten, 
       Astrophysics and Space Science Proceedings, Springer-Verlag, 
       Heidelberg, Berlin, 2009.} \vspace*{-5ex} \par
 \def\lastand{\ifnum\value{@inst}=2\relax
                 \unskip{} \andname\
              \else
                 \unskip \lastandname\
              \fi}%
 \def\and{\stepcounter{@auth}\relax
          \ifnum\value{@auth}=\value{@inst}%
             \lastand
          \else
             \unskip,
          \fi}%
  \raggedright
 {\Large \bfseries\boldmath
  \pretolerance=10000
  \let\\=\newline
  \raggedright
  \hyphenpenalty \@M
  \interlinepenalty \@M
  \if@numart
     \chap@hangfrom{}
  \else
     \chap@hangfrom{\thechapter\thechapterend\hskip\betweenumberspace}
  \fi
  \ignorespaces
  \@title \par}\vskip .8cm
\if!\@subtitle!\else {\large \bfseries\boldmath
  \vskip -.65cm
  \pretolerance=10000
  \@subtitle \par}\vskip .8cm\fi
 \setbox0=\vbox{\setcounter{@auth}{1}\def\and{\stepcounter{@auth}}%
 \def\thanks##1{}\@author}%
 \global\value{@inst}=\value{@auth}%
 \global\value{auco}=\value{@auth}%
 \setcounter{@auth}{1}%
{\lineskip .5em
\noindent\ignorespaces
\@author\vskip.35cm}
 {\small\institutename\par}
 \ifdim\pagetotal>157\p@
     \vskip 11\p@
 \else
     \@tempdima=168\p@\advance\@tempdima by-\pagetotal
     \vskip\@tempdima
 \fi
}
\makeatother     
\fi

\title*{Theoretical Models of Sunspot Structure and Dynamics}

\author{J. H. Thomas}
\authorindex{Thomas, J. H.}

\institute{Department of Mechanical Engineering and Department of 
Physics \& Astronomy, University of Rochester, USA}

\maketitle

\setcounter{footnote}{0}  

\begin{abstract} 

Recent progress in theoretical modeling of a sunspot is reviewed. The 
observed properties of umbral dots are well reproduced by realistic simulations 
of magnetoconvection in a vertical, monolithic magnetic field. To understand 
the penumbra, it is useful to distinguish between the inner penumbra, 
dominated by bright filaments containing slender dark cores, and the outer 
penumbra, made up of dark and bright filaments of comparable width with 
corresponding magnetic fields differing in inclination by some 30 degrees 
and strong Evershed flows in the dark filaments along nearly horizontal or 
downward-plunging magnetic fields. The role of magnetic flux pumping in 
submerging magnetic flux in the outer penumbra is examined through 
numerical experiments, and different geometric models of the penumbral 
magnetic field are discussed in the light of high-resolution observations. 
Recent, realistic numerical MHD simulations of an entire sunspot have 
succeeded in reproducing the salient features of the convective pattern 
in the umbra and the inner penumbra. The siphon-flow mechanism still
provides the best explanation of the Evershed flow, particularly in the 
outer penumbra where it often consists of cool, supersonic downflows.

\end{abstract}

\section{Introduction}      \label{thomas-sec:introduction}

Understanding the structure and dynamics of a sunspot poses a formidable
challenge to magnetohydrodynamic theory. The marvelous details revealed 
in high-resolution observations of sunspots have shown how very complex a 
sunspot is, but have also stimulated real progress in theoretical modeling. 

Here I review recent advances on some important theoretical issues
concerning sunspots, including the following questions. Is the overall
near-surface structure of a sunspot best described as a monolithic 
(but inhomogeneous) magnetic flux tube or as a cluster of individual
flux tubes? What is the nature of magnetoconvection in a sunspot, and how
does it produce the umbral dots and the filamentary intensity pattern 
in the penumbra? What causes the complicated interlocking-comb configuration
of the magnetic field in the penumbra? How do we explain the significant 
differences between the inner and outer penumbra? What causes the Evershed
flow in the penumbra? How do the outflows along the dark penumbral cores in
bright filaments in the inner penumbra relate to the stronger and 
downward plunging Evershed flows in the outer penumbra? 

This review is of necessity selective, and some important topics will not 
be discussed at all (for example, sunspot seismology, which is well
covered by Rajaguru and Hanasoge in this volume).
For a broader coverage of both theory and observations of sunspots, see the 
recent book by Thomas \& Weiss (2008) and the reviews by Solanki (2003) 
and Thomas \& Weiss (2004).

\section{Umbral magnetoconvection}	\label{thomas-sec:umbra}

In a broad sense, there are two competing models of the structure of a sunspot
below the solar surface: a monolithic, but inhomogeneous, magnetic 
flux tube, or a tight cluster of smaller flux tubes separated by field-free 
plasma (Parker 1979). One way in which we might distinguish between these 
two models is to examine the form of convective energy transport in the umbra, 
and in particular the mechanism that produces the bright umbral dots. 

In the monolithic model, the umbral dots are thought to correspond to slender,
hot, rising plumes that form within the ambient magnetic field and 
penetrate into the stable surface layer, spreading horizontally and sweeping
magnetic flux aside (flux expulsion), thereby producing a small, bright
region with a weakened magnetic field. This picture is supported by several
idealized model calculations involving both Boussinesq and fully compressible
magnetoconvection (see the reviews by Proctor 2005 and Thomas \& Weiss 2008). 
In the cluster model, convection is imagined to be effectively suppressed 
in the magnetic flux tubes but unimpeded in the nearly field-free regions 
around them, where the convection penetrates upward into the visible layers 
to form bright regions. In that case, however, we might reasonably expect to 
see a bright network enclosing dark features, rather than the observed pattern 
of bright, isolated umbral dots on a dark background (e.g., Knobloch \& Weiss 1984). 
The essential differences between the monolith and cluster models are that
in the cluster model the weak-field gaps are permanent and are connected
to the field-free plasma surrounding the sunspot, whereas in the monolithic 
model the gaps are temporary and are embedded within the overall flux
tube, isolated from the surroundings of the spot. 

\begin{figure}
\sidecaption
\includegraphics[height=80mm]{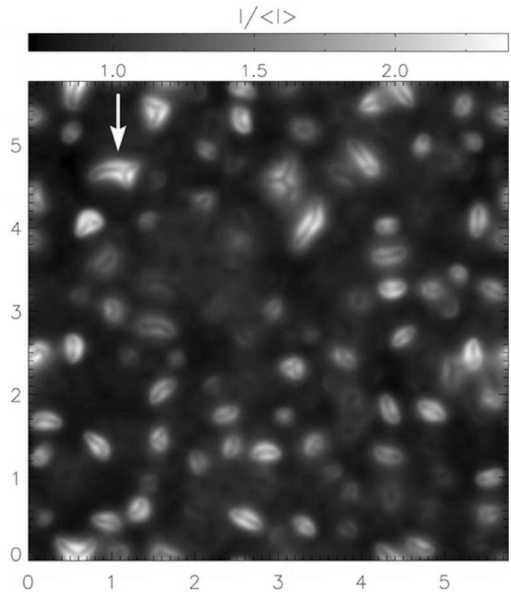}
\caption[]{\label{thomas-fig:umbral_dots} The pattern of vertically 
emerging surface intensity in a realistic numerical simulation of umbral 
magnetoconvection. (From Sch\"ussler \& V\"ogler 2006.)}
\end{figure}

Recently, Sch\"ussler \& V\"ogler (2006) carried out realistic numerical 
simulations of umbral magnetoconvection in the context of the monolithic 
model, assuming an initially uniform vertical magnetic field. They study 
three-dimensional compressible magnetoconvection within a realistic 
representation of an umbral atmosphere, including partial ionization 
effects and radiative transfer. Their model reproduces all of the principal
observed features of umbral dots (see, e.g., Bharti, Jain \& Jaaffrey 2007).
The results show an irregular pattern of slender, isolated plumes of width 
200--300 km and lifetime around 30 min. An individual plume achieves a peak 
upward velocity of about 3 km s$^{-1}$ before 
decelerating (by buoyancy braking) and spreading laterally as it meets the 
stable surface layer, greatly reducing the local magnetic field strength. 
Figure~\ref{thomas-fig:umbral_dots} shows a snapshot of the emerging intensity 
at the surface corresponding to optical depth $\tau_{500} = 1$ (which 
is elevated above the rising plumes). Note that the plumes are generally oval 
rather than circular in shape, and they have dark streaks along their major 
axes. These dark streaks are absorption features caused by the local increase 
of density and pressure associated with buoyancy braking of the plumes
(cf. Section~\ref{thomas-sec:simulations}). The dark streaks have been seen
in Hinode observations (Bharti et al.\ 2009). 

While the results of Sch\"ussler \& V\"ogler do not necessarily rule out the 
cluster model, they do provide strong support for the monolithic model, in the 
sense that they show that umbral dots arise naturally as a consequence of 
magnetoconvection in a space-filling, vertical magnetic field. The magnetic 
flux is partially expelled from the plume regions to allow convective motions 
to occur, but these regions are not entirely field free and, more importantly, 
they are isolated within the overall flux bundle and not in contact with 
field-free plasma below, as they would be in the cluster model.  


\section{The inner and outer penumbra}	\label{thomas-sec:inner-outer}

In understanding the structure of the penumbra, it is useful to distinguish between
the {\it inner} and the {\it outer} penumbra (Brummell et al.\ 2008). The boundary 
between them is somewhat arbitrary, but it may be conveniently defined as the line 
separating inward-moving and outward-moving grains in the bright filaments, lying 
at about 60\% of the radial distance between the inner and outer edges of the penumbra
and dividing the penumbra into roughly equal surface areas (Sobotka, Brandt \& 
Simon 1999; Sobotka \& S\"utterlin 2001; M\'arquez, S\'anchez Almeida \& Bonet 2006). 
This pattern may be understood as a transition from isolated, vertical convective 
plumes in the umbra to elongated, roll-like convective structures in the outer 
penumbra, as a consequence of the increasing inclination (to the local vertical) 
of the magnetic field. The moving bright grains are then traveling patterns of 
magnetoconvection in an inclined magnetic field, with the motion switching from 
inward to outward at some critical inclination angle of the magnetic field. 

The inner penumbra is dominated by bright filaments containing slender dark cores
(Scharmer et al.\ 2002; Langhans et al.\ 2007) and has relatively small 
azimuthal variations in the inclination of the magnetic field.  The field in 
a dark core is slightly more inclined than the field in its bright surroundings, 
by some 4--10$\deg$. A dark core typically originates at a bright feature near 
the umbra, where there is an upflow that bends over into an outflow along the 
inclined magnetic field in the core. 

The outer penumbra, on the other hand, is made up of dark and bright filaments of 
comparable width, with corresponding magnetic fields differing significantly in 
inclination, by 20--30$\deg$ or more, the more horizontal field being in
the dark filaments. The Evershed flow is stronger in the outer penumbra and is
generally concentrated in the dark filaments, along nearly horizontal and often 
downward-plunging magnetic fields, with the flow velocity and the magnetic field 
well aligned all along the filament. One of the most intriguing features of the 
outer penumbra is the presence of ``returning'' magnetic flux, that is, field lines 
with inclinations greater than $90\deg$ that plunge back below the solar 
surface. There is now overwhelming observational evidence for a substantial 
amount of returning magnetic flux in the outer penumbra, in several high-resolution 
polarimetric studies based on different inversion schemes (e.g., Westendorp Plaza 
et al.\ 2001; Bellot Rubio, Balthasar \& Collados 2004; Borrero et al.\ 2004; 
Langhans et al.\ 2005; Ichimoto et al.\ 2007, 2009; Beck 2008; Jur\u{c}\'ak \& 
Bellot Rubio 2008).

The outer edge of the penumbra is quite ragged, with prominent dark filaments 
protruding into the surrounding granulation. The proper motions of granules in 
the moat surrounding a spot show convergence along radial lines extending outward 
from the protruding dark filaments (Hagenaar \& Shine 2005), providing evidence 
for submerged magnetic flux extending outward from the spot. This submerged magnetic 
field is presumably held down, in opposition to its inherent buoyancy, by magnetic 
flux pumping, as described in the next section.

\section{The formation and maintenance of the penumbra}  \label{thomas-sec:formation}

One of the important challenges for sunspot theory is to explain how the 
filamentary penumbra forms and its magnetic field acquires the observed
interlocking comb structure with downward-plunging field lines in the outer
penumbra, and how this structure is maintained. Eventually this whole process
may be amenable to direct numerical simulation (see 
Section~\ref{thomas-sec:simulations} below), but for now we can only speculate 
based on less ambitious models of specific aspects of the process. 

The following scenario has been suggested for the formation of a fully
fledged sunspot with a penumbra (Thomas et al.\ 2002; Weiss et al.\ 2004).
The development of a solar active region begins with the emergence of a 
fragmented magnetic flux tube into the photosphere. The emergent flux is 
quickly concentrated into small, intense magnetic elements which can
accumulate in the lanes between granules and mesogranules to form small
pores. Some of these pores and magnetic elements may then coalesce to 
form a sunspot. Simple models show that, as a growing pore accumulates 
more magnetic flux, the inclination (to the local vertical) of the magnetic 
field at its outer boundary increases until it reaches a critical value, 
whereupon a convectively driven fluting instability sets in and a penumbra 
forms. The fluting of the magnetic field near the outer boundary of the 
sunspot's flux tube brings the more horizontal spokes of field into 
greater contact with the granulation layer in the surroundings, and
then downward magnetic pumping of this flux by the granular convection 
further depresses this magnetic field to form the ``returning'' magnetic 
fields (inclination greater than 90$\deg$) seen in the outer penumbra.
The transition between a pore and a sunspot shows hysteresis, in the sense
that the largest pores are bigger than the smallest sunspots; this may 
be explained by the flux-pumping mechanism, which can keep the fields
in the dark filaments submerged even when the total flux in a decaying spot
is less than that at which the transition from pore to spot occurred.

We have demonstrated the efficacy of the process of magnetic flux pumping 
by granular convection
through a series of idealized numerical experiments (Thomas et al.\ 2002;
Weiss et al.\ 2004), most recently for a more realistic, arched magnetic 
field configuration that accounts more accurately for the magnetic 
curvature forces (in addition to the buoyancy forces) opposing the downward 
pumping (Brummell et al.\ 2008). We solve the equations governing 
three dimensional, fully compressible, nonlinear magnetoconvection 
in a rectangular box, consisting of two layers: an upper, superadiabatic
layer of vigorous convection representing the granulation layer, and a lower, 
marginally stable or weakly superadiabatic layer representing the rest of 
the convection zone. 
The simulation is run without a magnetic field until a statistically 
steady state is reached, and then a strong magnetic field is introduced, 
in the form of a purely poloidal ($x$--$z$), double arched magnetic field,
and the gas density is adjusted to maintain pressure equilibrium. 
The calculation proceeds and we examine the effect of the convection 
in redistributing the magnetic flux. 

\begin{figure}
\centering
\includegraphics[width=\textwidth]{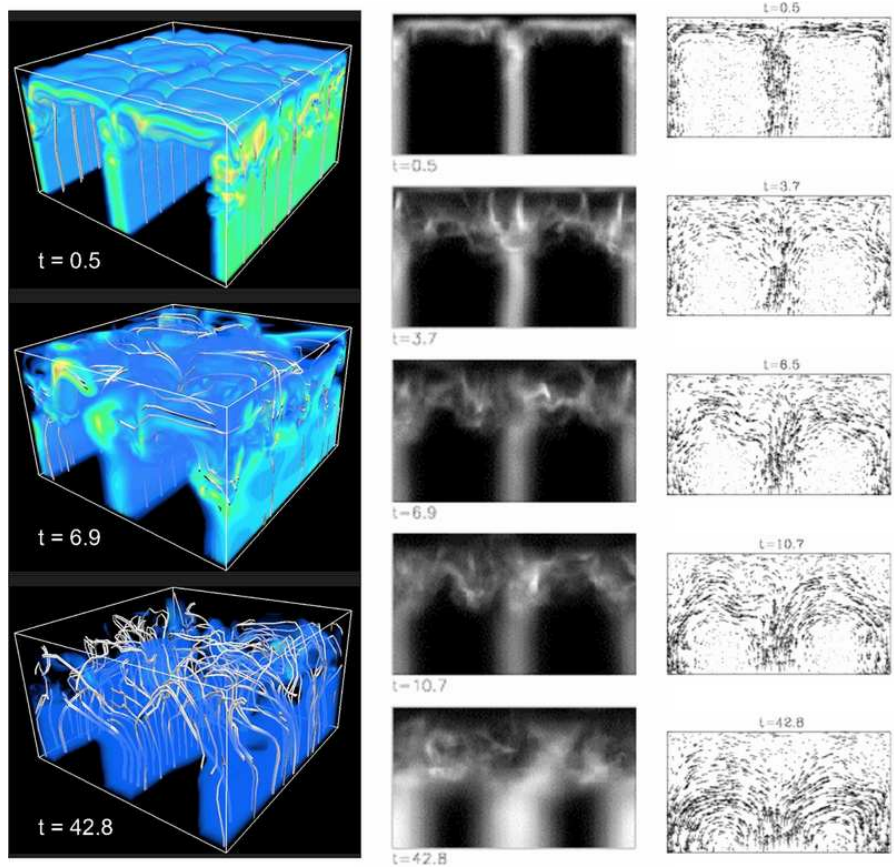}
\caption[]{\label{thomas-fig:flux_pumping} Numerical simulation of downward
magnetic flux pumping of penumbral magnetic fields by granular convection
in the surroundings. The panels show volume renderings of magnetic energy 
density (left), $x$-averaged magnetic energy density (middle), and the 
$x$-averaged vector magnetic field at different times during the run. 
(Here $x$ is the direction perpendicular to the page.) The initial 
arched magnetic field configuration is still distinctly visible in the 
uppermost plots at $t=0.5$. (From Brummell et al.\ 2008.)}
\end{figure}

Figure~\ref{thomas-fig:flux_pumping}
shows the state of the magnetic field shortly after it was introduced
(scaled time $t=0.5$) and at a few later stages, the last stage ($t=42.8$) 
being after a new quasi-steady statistical state has been reached. Here 
we see that a significant fraction of the large-scale magnetic field is 
pumped rapidly downward out of the upper granulation layer and concentrated
mostly in the upper part of the lower, more quiescent convective layer. 
These new numerical experiments demonstrate that the downward pumping
by turbulent granular convection is indeed able to overcome the combined 
effects of the magnetic buoyancy force and the curvature force due to 
magnetic tension, and thus to submerge much of the initial, nearly
horizontal magnetic flux beneath the granulation layer, as we propose
in the scenario presented above.

\section{The magnetic field configuration in the penumbra} \label{thomas-sec:mag_pen}

\begin{figure}
\centering
\includegraphics[height=55mm]{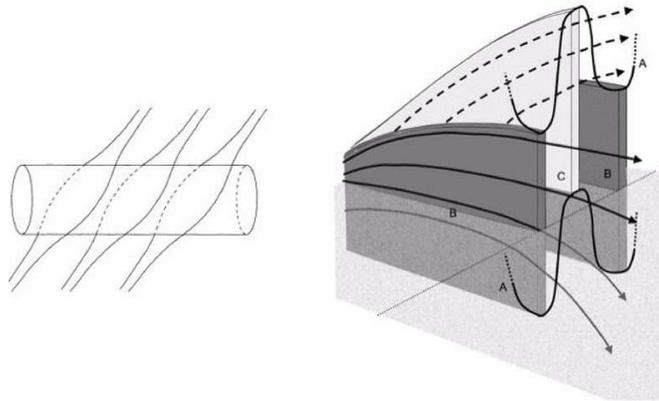}
\caption[]{\label{thomas-fig:sketches} Two simple models of the penumbral 
magnetic field configuration. Left panel: Sketch of the magnetic field
configuration in the ``uncombed'' penumbral model of Solanki \& Montavon (1993),
with an ambient magnetic field wrapping around a thin horizontal flux tube
(dark filament). Right panel: Schematic diagram of the ``interleaved sheet'' 
model of the outer penumbra (Brummell et al.\ 2008.), with a fluted magnetopause 
(A) and slabs of nearly horizontal magnetic field (B, dark filaments) extending
downward to some depth below the surface and separated by a slab of less steeply 
inclined magnetic field (C, bright filament).}
\end{figure}

Here we consider some geometric models that have been proposed for the 
observed interlocking-comb structure of the penumbral magnetic field. 
The scenario described in Section~\ref{thomas-sec:formation} for the formation 
of the penumbra and the returning flux tubes through flux pumping
leads us to a magnetic field configuration in the outer penumbra roughly as
depicted in the right-hand panel of Figure~\ref{thomas-fig:sketches} 
(Thomas et al.\ 2006; Brummell et al.\ 2008). This configuration, which we 
might describe as an ``interleaved sheet'' model, has vertical sheets of nearly 
horizontal magnetic field (dark filaments) interleaved between sheets of more 
vertical magnetic field (bright filaments). In this picture, the sheets of 
horizontal field extend downward below the visible surface to a depth of, say, 
5 Mm. (A simple estimate gives the depth of penetration equal to one-quarter 
of the width of the penumbra: Brummell et al.\ 2008).

Another geometric model, with a longer history, is the ``uncombed'' penumbral 
model\footnote{Sometimes the term ``uncombed'' is used more generally to describe 
the observed penumbral field configuration, but here I use the term specifically 
to represent the geometric model proposed by Solanki \& Montavon (1993).}
of Solanki \& Montavon (1993), depicted in the left-hand panel of 
Figure~\ref{thomas-fig:sketches}.
In this model the more horizontal component of the penumbral magnetic field is 
represented by horizontal magnetic flux tubes, of nearly circular 
cross-section, embedded in a more vertical background magnetic field that wraps 
around these tubes. Scharmer \& Spruit (2006) pointed out that the magnetic tension
forces in the background magnetic field will tend to compress a circular flux tube 
in the horizontal direction, causing it to expand upward at the top and downward 
at the bottom, perhaps indefinitely. Borrero, Rempel \& Solanki (2006) then argued 
that buoyancy forces will halt this squeezing process, leaving a flux tube of tall, 
narrow cross-section. If the vertical elongation of the flux tube is significant,
the configuration begins to look much like the interleaved sheet model depicted 
in the right-hand panel of Figure~\ref{thomas-fig:sketches}, and these two
models are then not very different.

\begin{figure}
\centering
\includegraphics[height=35mm]{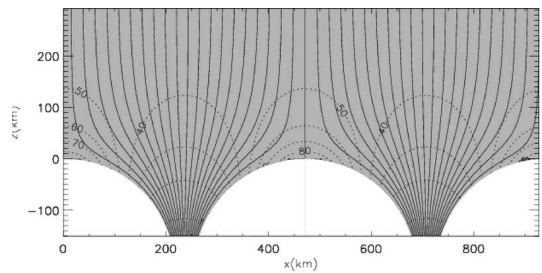}
\caption[]{\label{thomas-fig:gappy} The potential magnetic field configuration
in the ``gappy penumbra'' model of Spruit \& Scharmer (2006). Shown here are
the magnetic field lines (solid lines) projected onto a vertical ($x$--$z$) plane 
perpendicular the axis ($y$-axis) of a penumbral filament, along with contours 
(dotted lines) of constant inclination of the field in the $y$--$z$ plane.}
\end{figure}

A quite different model of the penumbral magnetic field, the ``gappy penumbra'' model 
of Spruit \& Scharmer (2006; Scharmer \& Spruit 2006), is based on the 
cluster model of a sunspot. It postulates field-free, radially aligned gaps in 
the magnetic field below the visible surface of the penumbra, protruding into 
a potential magnetic field configuration. The gaps are assumed to extend 
indefinitely downward, allowing the field-free convection in the gaps to carry 
the bulk of the upward heat flux in the penumbra. 
Figure~\ref{thomas-fig:gappy} shows the proposed magnetic field configuration.
The gaps themselves represent the bright penumbral filaments, 
while the intervening regions of strong magnetic field represent the dark filaments. 
As can be seen from the contours of constant inclination in 
Figure~\ref{thomas-fig:gappy}, the magnetic field is more nearly horizontal 
above the bright filaments (the gaps) and more nearly vertical (here 45\deg) 
above the dark filaments. However, this magnetic field configuration is in 
direct contradiction with numerous observations that show that the 
field is more horizontal in the dark filaments (e.g., Rimmele 1995; Stanchfield, 
Thomas \& Lites 1997; Westendorp Plaza et al.\ 2001; Langhans et al.\ 2005),
including very recent spectropolarimetric observations from Hinode by
Jur\u{c}\'ak \& Bellot Rubio (2008) and by Borrero \& Solanki (2008). The 
last authors also examined the vertical stratification of magnetic field 
strength in the penumbra and found that it is inconsistent with the existence 
of regions void of magnetic field at or just below the $\tau_{500} = 1$ level.
While the gappy penumbra model itself contains no flows, Spruit \& Scharmer 
suggest that the Evershed flow occurs along the (very restricted) region of 
nearly horizontal field just above the center of the gap. At least in the 
outer penumbra, this is in conflict with numerous observations that show that 
the flow is concentrated in the dark filaments. It seems, then, that the 
gappy penumbra is incompatible with observations. 

Spruit \& Scharmer (2006) also suggested that the observed narrow dark cores 
running along the center of bright filaments in the inner penumbra can be 
understood as an effect of the increased opacity due to increased gas pressure 
in the field-free gaps. This important suggestion seems to be basically correct, 
although the field-free gaps are not necessary: dark cores also form as opacity
effects in the case of magnetoconvection in a strong-field region, as shown in 
the simulations of umbral dots discussed in Section~\ref{thomas-sec:umbra} above 
and in the simulations of penumbral filaments discussed in the next section.

\section{Numerical simulations of a sunspot}	\label{thomas-sec:simulations}

Any attempt to perform a direct numerical MHD simulation of an entire sunspot 
faces serious computational difficulties: the simulation must represent a 
very large structure while still resolving fine-scale features and even 
smaller scale diffusive effects; it must cope with a wide range of values
of the Alfv\'en speed and plasma beta; and the computation must be carried 
out long enough to reach a relaxed, quasi-steady state. In spite of these
formidable problems, there have been very recently impressive attempts by 
two groups to model an entire sunspot by direct, realistic simulations 
including radiative transfer (Heinemann, Nordlund, Scharmer \& Spruit 2007; 
Rempel, Sch\"ussler \& Kn\"olker 2009). These efforts are surely just the 
beginning of a new and fruitful approach to sunspot theory. 

\begin{figure}
\centering
\includegraphics[height=95mm]{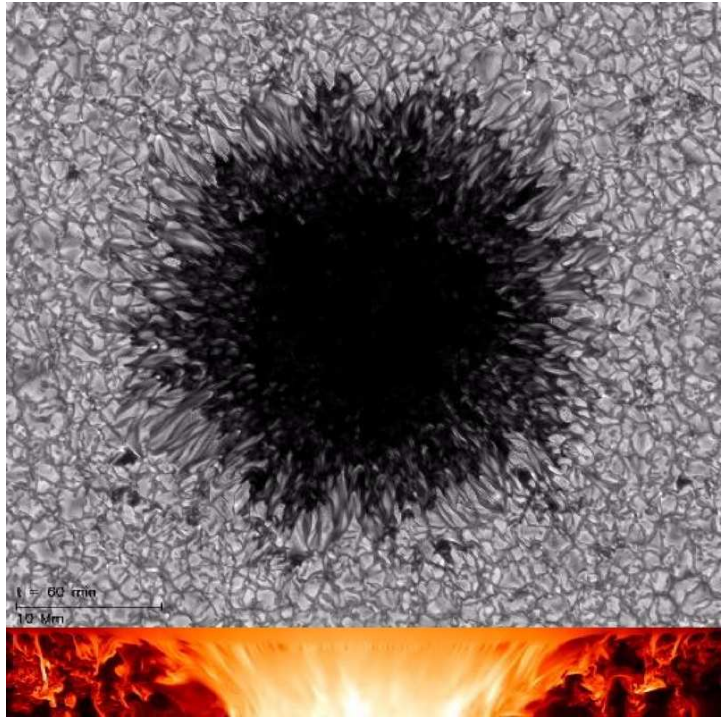}
\caption[]{\label{thomas-fig:Rempel2} Numerical simulation of a circular sunspot, 
following the same method as in the slab model of Rempel et al.\ (2009). Shown
here are (above) a snapshot of the surface intensity and (below) the corresponding
values of $|B|^{1/2}$ on a vertical slice through the center of the spot,
depicted on a color scale. (Courtesy of Matthias Rempel.)}
\end{figure}

Both groups model a large section of a sunspot in a rectangular
box. They each introduce a two-dimensional, vertically spreading, initial 
magnetic field into a state of fully developed nonmagnetic convection 
representing the upper convection zone and a stable atmospheric layer above it. 
The calculations continue for several hours of real (solar) time, through
a dynamic adjustment phase, until a quasi-steady state is attained. 
The results show the formation of filamentary structures resembling those
in the inner penumbra of a real sunspot, including bright filaments 
containing central dark cores.

In very recent work, Rempel et al.\ have extended their simulations to 
model an entire circular sunspot within a rectangular box. 
Figure~\ref{thomas-fig:Rempel2} shows a snapshot of the surface intensity 
pattern and magnetic field in this beautiful simulation.

\begin{figure}
\centering
\includegraphics[height=5cm]{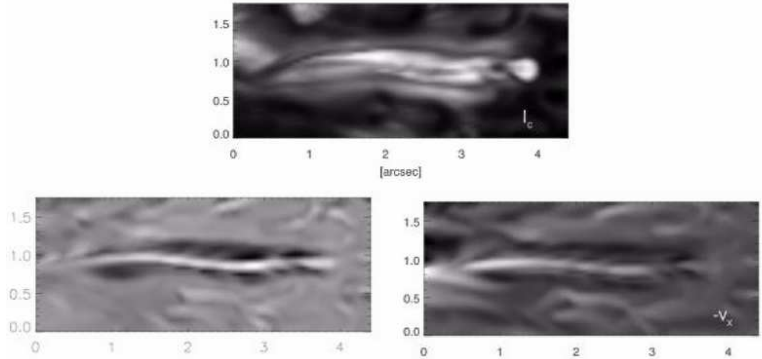}
\caption[]{\label{thomas-fig:Rempel1} Enlarged view of a single bright
penumbral filament produced in the simulation of Rempel et al.\ (2009).
The umbra lies to the right of this filament.
The upper panel shows a surface continuum intensity image at wavelength
630 nm, and the lower panels show vertical velocity $v_z$ (left) and horizontal
velocity $v_x$ (right), where the $x$-axis is parallel to the bottom of
the panels.}
\end{figure}

The simulations reproduce most of the important features of the 
bright penumbral filaments found in the inner penumbra. 
Figure~\ref{thomas-fig:Rempel1} shows a blowup of a single bright 
penumbral filament produced in the rectangular sunspot simulation of 
Rempel et al.\ (2009). The continuum intensity pattern shows an elongated 
bright filament with a dark central core and a bright ``head'', which
migrates inward toward the umbra during the lifetime of the filament. 
The dark core is produced as an opacity effect due to buoyancy braking 
of the upflow, much as in the simulations of umbral dots discussed in 
Section~\ref{thomas-sec:umbra}. The magnetic field (not shown here) is 
weaker and more inclined in the filament than in its immediate surroundings.

The pattern of vertical velocity shows roll-like convection along the 
filament with an inclined upflow along the central axis of the filament 
(i.e., along the dark core) and inclined downflows (return flows) along 
the sides of the filament. Correspondingly, the radial ($x$) component 
of the velocity is outward along the axis of the filament, with a peak 
value of about 2 km s$^{-1}$, and inward and along the sides of the filament.
The return flow is in regions with stronger and less inclined magnetic
field, so the horizontal component is smaller in magnitude than that 
in the outflow at the same optical depth; as a result, the radial inflows
and outflows do not cancel when averaged in the $y$-direction, but instead
show an average outflow of about 1 km s$^{-1}$. 
  
The simulated penumbral filaments are slender structures with a width
of a few hundred km and a depth of about 2 Mm. The filaments form and
remain embedded within an overall region of strong magnetic field, and
they are well isolated from the field-free convection beneath the
penumbra.  As Rempel et al.\ emphasize, these weak-field ``gaps''
formed within the overall magnetic field by the convection are
fundamentally different from those proposed in the ``gappy penumbra''
model of Spruit \& Scharmer, which are protrusions of the exterior
field-free plasma into the penumbra as envisioned in the cluster model
(see Section~\ref{thomas-sec:mag_pen}).  At a fundamental level, the
simulations discussed in this section are based on the monolithic
model and they support that model by producing results that match
observations of the inner penumbra; they lend no support for the
``gappy penumbra''.

Rempel et al.\ also point out that their results do not support the ``moving 
tube'' model of Schlichenmaier, Jahn \& Schmidt (1998): vertical heat transport 
takes place all along the simulated filaments, not just along separate, thin 
flux tubes, and the movement of the filament ``heads'' inward toward the penumbra 
is due to a propagation of the magnetoconvective pattern rather than the 
bodily motion of an individual thin flux tube.

In both simulations discussed above, the overall extension of the penumbra
is rather small and the inclination of the magnetic field in the outer
part of the penumbra is generally much less than that found in a real
sunspot. Thus, as both groups admit, the simulations so far seem to 
reproduce only the inner penumbra. One reason for this is that the periodic 
boundary conditions employed effectively place another sunspot of the same 
magnetic polarity nearby, on either side of the simulated spot. This hinders 
the formation of nearly horizontal fields in the outer penumbra. (Indeed, 
observations show that sunspots often do not form a penumbra in a sector near 
another spot of the same polarity.) This could be remedied, for example, 
by using periodic boundary conditions like those of Brummell et al.\ (2008), 
which produce a row of spots of alternating polarity 
(see Fig.~\ref{thomas-fig:flux_pumping}).

\section{The Evershed flow}	\label{thomas-sec:Evershed}

Since the occasion for this meeting is the centennial of John Evershed's 
discovery, it seems appropriate to close with some remarks about theoretical 
interpretations of the Evershed flow. 
The flow occurs along arched, elevated flow channels. Recent results from 
Hinode support this picture. Ichimoto et al.\ (2007) find that the Evershed 
downflows in the outer penumbra have the flow velocity vector and magnetic 
field vector well aligned, at an angle of about $30\deg$ to the solar 
surface. Jur\u{c}\'ak \& Bellot Rubio (2008) find that the average 
inclination of the magnetic field associated with the Evershed flow channels 
increases from $85\deg$ to $105\deg$ in going from the middle to 
the outer penumbra, quite consistent with the earlier results of 
Langhans et al.\ (2005) from the Swedish Solar Telescope. 

The arched nature of the flow channels and the strong, often supersonic,
field-aligned downflows in the outer penumbra are well reproduced 
in the siphon flow model (e.g., Montesinos \& Thomas 1997). The 
``moving tube'' model of Schlichenmaier, Jahn \& Schmidt (1998) does 
not produce this configuration: it has no returning flux or
downflow, but instead has all of the flow continuing radially outward 
along the elevated magnetic canopy. Schlichenmaier (2002) did find
a class of super-Alfv\'enic, serpentine solutions for his model,
which do have downflows along a returning flux tube, but these flows 
are unphysical: the very high flow speeds are an artifact of the
outer boundary condition, and moreover the flow configuration itself
is gravitationally unstable (Thomas 2005) and hence will not occur. 
(This instability seems to have been ignored by some, however, and 
the serpentine solutions continue to be invoked as a possible 
explanation of the Evershed flow: e.g., Schlichenmaier, M\"uller \& 
Beck 2007; Sainz Dalda \& Bellot Rubio 2008.)

The numerical simulations discussed in the previous section produce an 
outward horizontal velocity component of 1--2 km s$^{-1}$ along the axis of a
filament (see Fig. 6), which might explain the radial outflows seen in 
the dark cores in the inner penumbra, although it is not clear why the 
associated inflows along the sides of the core are not observed. However, 
the simulations do not offer a complete explanation of the Evershed flow,
as claimed by Scharmer, Nordlund \& Heinemann (2008). In the simulations,
the peak outward velocity is only about 2 km s$^{-1}$ and the outward speed 
averaged over a few filaments is only about 1 km s$^{-1}$, considerably less 
than what is observed in the outer penumbra. The simulations do not come 
close to producing the supersonic flow speeds of 7--16 km s$^{-1}$,
aligned with downward-plunging returning flux tubes, 
that are observed in dark filaments in the outer penumbra (e.g., Westendorp 
Plaza et al.\ 2001; del Toro Iniesta, Bellot Rubio \& Collados 2001; 
Penn et al.\ 2003).

The supersonic, cool Evershed downflows are an inherent feature of the 
siphon-flow model (Montesinos \& Thomas 1997). Siphon flows still provide 
the best description of the Evershed flows in the outer penumbra, although
the flows computed so far have all been steady state and thus do not
reproduce the transient nature of flows. A thin-flux-tube
model combining the best features of the siphon-flow model (arched, 
returning flux tubes, cool supersonic downflows) and the moving-tube 
model (transient flows, heating at the inner footpoint) would likely 
reproduce all of the salient features of the Evershed flow. 

In a broad sense the Evershed flow must fundamentally be a convective 
phenomenon. Even in the models based on thin flux tubes -- the moving
tube model or the siphon-flow model -- the flow is driven by a pressure 
difference along the tube produced by some combination of local heating
(producing an increase in gas pressure) or convective collapse
(producing a decrease in gas pressure), and the returning flux is 
produced by turbulent convective pumping.  As computing capabilities 
increase and the numerical simulations succeed in resolving all 
aspects of the convection in a sunspot and its immediate 
surroundings, we can expect the Evershed flow and the returning flux
tubes to be a natural outcome. 

\section{Conclusions}	\label{thomas-sec:Conclusions}

The principal conclusions of this review are the following: 

\begin{itemize}

\item The observed properties of umbral dots are well explained by realistic 
simulations of magnetoconvection in a vertical, monolithic magnetic field;
there is no need to invoke a cluster model.

\item There are significant differences between the inner and outer penumbra,
and it is useful to distinguish between them.

\item Downward pumping of magnetic flux by turbulent granular convection
offers a plausible mechanism for producing the returning magnetic flux in
the outer penumbra. 

\item The ``uncombed'' and ``interleaved sheet'' models of the penumbral
magnetic field configuration are actually quite similar, in view
of the squeezing effect on the circular flux tubes in the uncombed
model.

\item The ``gappy penumbra'' model for the penumbral magnetic field configuration
is not in accord with observations. 

\item Recent realistic simulations of an entire sunspot have succeeded in 
reproducing the structure of the inner penumbra. However, they do not 
reproduce the structure of the outer penumbra, with its horizontal and 
returning magnetic fields and fast (supersonic) Evershed flows along
arched channels. 

\item Bright penumbral filaments in the inner penumbra are well reproduced
in these simulations, as roll-like convection (not interchange convection).
Magnetic flux is partially expelled by the convective plumes, but the 
resulting ``gaps'' are not in contact with the exterior plasma and hence are 
fundamentally different from the gaps in the ``gappy penumbra'' model.
The simulations reproduce the central dark cores in the bright filaments,
as an opacity effect due to buoyancy braking of the plumes, and the 
outflows seen in these cores.

\item The siphon-flow model still provides the best description of the 
Evershed flow in the outer penumbra. The moving-tube model describes
the transient nature of the Evershed flow but fails to produce returning 
flux tubes and downflows. A thin-flux-tube model combining the best features 
of these two models is suggested.

\end{itemize}


\begin{acknowledgement}
I thank Siraj Hasan for making it possible for me to attend this meeting, and
Matthias Rempel for providing results and figures prior to their publication. 
I also thank my collaborators Nic Brummell, Steve Tobias, and Nigel Weiss, 
with special thanks to Nigel Weiss for many discussions of the topics and 
issues covered in this review.
\end{acknowledgement}

\begin{small}

\end{small}

\end{document}